\pdfoutput=1
\documentclass[preprint]{elsarticle}
\usepackage{amsmath}
\usepackage{amssymb}
\usepackage[T1]{fontenc}
\usepackage{pdfpages} 

\begin{document}

\begin{frontmatter}

\title{Mathematical model of livestock and wildlife: Predation and competition under environmental disturbances}

\author[fiestin]{M. F. Laguna\corref{cor1}}
\ead{lagunaf@cab.cnea.gov.ar}
\cortext[cor1]{Corresponding author}
\author[fiestin,ib]{G. Abramson}
\ead{abramson@cab.cnea.gov.ar}
\author[fiestin,ib]{M. N. Kuperman}
\ead{kuperman@cab.cnea.gov.ar}
\author[iidivca]{J. L. Lanata}
\ead{jllanata@conicet.gov.ar}
\author[fb,rio]{J. A. Monjeau}
\ead{amonjeau@fundacionbariloche.org.ar}

\address[fiestin]{Centro At\'{o}mico Bariloche and CONICET, R8402AGP Bariloche, Argentina}
\address[ib]{Instituto Balseiro, R8402AGP Bariloche, Argentina}
\address[iidivca]{Instituto de Investigaciones en Diversidad Cultural y Procesos de Cambio, CONICET, R8400AHL Bariloche, Argentina}
\address[fb]{Fundación Bariloche and CONICET, R8402AGP Bariloche, Argentina}
\address[rio]{Laboratório de Ecologia e Conservação de Populações, Departamento de Ecologia, Universidade Federal do Rio de Janeiro, Rio de Janeiro, Brasil}

\begin{abstract}
Inspired by real scenarios in Northern Patagonia, we analyze a mathematical model of a simple trophic web with two herbivores and one predator. 
The studied situations represent a common practice in the steppes of Argentine Patagonia, where livestock are raised in a semi-wild state, either on the open range or enclosed, coexisting with competitors and predators. In the present work, the competing herbivores represent sheep and guanacos, while the predator is associated with the puma. 
The proposed model combines the concepts of metapopulations and patches dynamics, and includes an explicit hierarchical competition between species, which affects their prospect to colonize an empty patch when having to compete with other species.
We perform numerical simulations of spatially extended metapopulations assemblages of the system, which allow us to incorporate the effects of habitat heterogeneity and destruction. The numerical results are compared with those obtained from mean field calculations.
We find that the model provides a good theoretical framework in several situations, including the control of the wild populations that the ranchers exert to different extent. Furthermore, the present formulation incorporates new terms in previously analyzed models, that help to reveal the important effects due to the heterogeneous nature of the system.
\end{abstract} 


\date{\today}

\end{frontmatter}


\section{Introduction}

The mathematical modeling of ecological interactions is an essential tool in predicting the behavior of complex systems across changing scenarios, such as those arising from climate change or environmental degradation. The literature abounds with examples of predator-prey models~\cite{swihart01,bascompte98,kondoh2003}, of intra- and inter-specific competition~\cite{nee92,tilman94,hanski1983}, of the relation between species richness and area size~\cite{rosenzweig1995,ovaskainen2003} and of habitat fragmentation \cite{hanski2000,ovaskainen2002,hanski2002}. However, considerable effort still needs to be made in the integration of all these mechanisms together. Our intention is to advance towards the modeling of trophic web complexity in successive approximations. In this paper we take a first step in this direction: modeling a predator-prey-competition system in environments subjected to disturbances. We present as a case study the dynamics of a simple trophic web. As a paradigm of a more complex ecosystem, we analyze here the case of a single predator and two competing preys in the Patagonian steppe. Specifically, we focus on two native species: puma (\emph{Puma concolor}, a carnivore) and guanaco (\emph{Lama guanicoe}, a camelid), and on sheep (\emph{Ovis aries}) as an introduced competitor to the native herbivore and further prey of the pumas. This system is the result of a long sequence of ecological and historical events that we briefly outline below.

The native mammalian fauna of Patagonia is composed of survivors of five main processes of extinction. One of the most relevant is known as the Great American Biotic Interchange (GABI) in which, upon the emergence of the Isthmus of Panama about 3 million years ago, the South American biota became connected with North America~\cite{patterson2012}.
The last main event occurred during the Quaternary glaciations, when climate change was combined with the arrival of humans for the first time in the evolutionary history of the continent~\cite{martin1984}. This cast of species, resistant to these natural shocks, was afterwards not exempt from threats to their survival. 

Pumas and guanacos, currently the two largest mammals in Patagonia, coexisted with humans for at least 13000 years with no evidence of shrinkage in their ranges of distribution until the twentieth century~\cite{bastourre2012,borrero1996,denegris2005}. There are records of a huge abundance of guanacos in sustainable coexistence with Tehuelche hunters, until shortly before the arrival of the European immigrants~\cite{musters2007,claraz1988}. 

In 1880-1890, as a consequence of what has been called the ``Conquest of the Desert'' or ``Wingka Malón'' in Argentina, the indigenous highly mobile hunters were almost exterminated or driven from their ancestral territories by the campaigns of the Argentine military army~\cite{delrio2005}. Large and continuous extensions of the Patagonian steppe were subdivided into private ranches by means of a gigantic grid of fences, and 95\% of it was devoted to sheep farming~\cite{marques2011}. The introduction of sheep significantly altered the ecological interactions of the Patagonian fauna and flora~\cite{monjeau1989}. Ranchers, descendants of Europeans, built a new niche where the puma, as a predator of sheep, and the guanaco, as competitor for forage, became part of the list of enemies of their productive interests and were therefore fought~\cite{marques2011}.

\subsection{The ecological context}

Let us describe the current ecological scenario in which these three characteristic players interact. There is evidence of competition between sheep and guanacos~\cite{nabte2013,marques2011}, mainly for forage and water. From a diet of 80 species of plants, they share 76~\cite{baldi2004}, so that sheep carrying capacity decreases when the number of guanacos increases. Under natural conditions, the guanaco can be considered a superior competitor to the sheep, and in fact it has been observed that the guanacos displace sheep from water sites (including artificial sources). However, human influence makes the density of guanacos decrease when the number of sheep increases. What needs to be understood is that there is no simple and direct competition between guanacos and sheep, but a competition between guanacos and ``livestock,'' a term that includes sheep, herder dogs and humans with their guns. Without these ``cultural bodyguards'' of barks, bullets and fences, a herd of guanacos would displace a flock of sheep. As the fields deteriorate from overgrazing and desertification the guanaco increases its competitive superiority over the sheep, since it is superbly adapted to situations of environmental harshness, especially water scarcity (e.g., it may drink seawater in case of extreme necessity). Therefore, the guanaco density tends to increase naturally as field productivity decreases. However, this natural process is usually offset by an increase in hunting pressure on the guanaco as environmental conditions worsen, because ranchers want to maximize scarce resources for production. Drought periods catalyze these socio-environmental crises, as the lack of rain works somewhat like a destroyer of carrying capacity for both wildlife and livestock. 

Predators (pumas and also foxes, \emph{Lycalopex culpaeus}) are also subject to permanent removal, because their extermination reduces production costs~\cite{marques2011}. The puma naturally hunted guanacos~\cite{novaro2000}, but since the introduction of sheep it has been dedicated almost completely to these last, as it is a prey that involves minimal exploration cost (i.e., the energy effort spent in searching for, pursuing and capturing prey) compared to the high energy cost of capturing the fast and elusive guanaco that co-evolved with them~\cite{Rau2002}.

Of course, as hunting eradicates the predators, populations of herbivores have no natural demographic controls and can grow exponentially, as has happened in the case of some nature reserves. Subsequently the uncontrolled growth of guanaco increases the competition with sheep and, if guanaco populations are confined and cannot migrate, this may also result in an overconsumption of fodder, excessive destruction of the habitat, and subsequent population collapse by starvation. Such a case has recently been documented in the nature reserve of Cabo Dos Bahías in Patagonia ~\cite{marques2011}. 

This interaction between the guanaco, puma and sheep is heavily influenced by management decisions concerning the pastures. For ranchers, the fauna is a production cost, the tolerance of which can be characterized in three paradigmatic scenarios:

\textbf{Low conflict scenario.} If the cost of the presence of wildlife is financially compensated by the government or by ecotourism activities related to wildlife watching~\cite{nabte2013}, the ranchers tolerate the presence of wildlife in coexistence with a livestock density that is not harmful to the ecosystem. If the field changes from a productive use to a conservative use, the wildlife and the flora recover very fast (the San Pablo de Valdés case, see ~\cite{nabte2010}).

\textbf{Medium conflict scenario.} In well-managed fields with adequate load, the carrying capacity is maintained in a healthy enough state to tolerate livestock and wildlife simultaneously. For this to happen, the field needs a large usable area. If the field changes from a productive to a conservative use, the wildlife recovers well, but not as fast as in the previous scenario, since growth is limited by the availability of resources (the case of Punta Buenos Aires, Península Valdés). In those cases where the area is very large, the recovery is very good despite the partial deterioration, because the lower load per hectare is offset by the amount of surface (the case of Torres del Paine National Park, Chile).

\textbf{High conflict scenario.} When the farmer depends exclusively on sheep production, conflict is high because the wildlife goes against their economic interests. If these are not compensated,  wildlife hunting increases and becomes much more intense as the field deteriorates. Wildlife is shifted to productively marginal sectors. The farmer prioritizes short-term income above sustaining long-term productivity of the field. This economic rationality creates a negative feedback loop: as productivity decreases, the farmer increases livestock density to try to sustain the same income; this leads to overgrazing and, in turn, to a further decline in productivity; the farmer then sets out to economically compensate this decline by loading the field even more. The result is a meltdown of the productive system and the abandonment of the field~\cite{marques2011}. The populations of native herbivores (although they get a rest from the exterminated predators and the eradicated sheep) fail to recover viable population levels because there are not enough resources to sustain them, energetically and bio-geo-chemically~\cite{flueck2011}. As a consequence of this frequent scenario, the density of guanacos, rheas and other herbivores has decreased considerably in Patagonia~\cite{marques2011}.

In this paper we study a mathematical model corresponding to a simplified instance of this ecosystem. Our purpose is to provide a theoretical framework in which different field observations and conceptual models can be formalized. The model has been kept intentionally simple at the present stage, in order to understand the basic behavior of the system in different scenarios, including the three just described. Subsequent mechanisms will be incorporated and reported elsewhere. 

In the next section the metapopulation model is presented, followed by the analysis of the main results obtained with the spatially explicit stochastic simulations and the mean field model. Further discussion and future directions are discussed in the closing section.

\section{Analysis}

\subsection{Metapopulation model}

As we said above, let us analyze a system composed of three characteristic species of the trophic web: two competing herbivores (sheep and guanacos), and a predator (puma). A particularly suited framework to capture the role of a structured habitat and of the hierarchical competition is that of metapopulation models~\cite{tilman97,tilman94}. A metapopulation model comprises a large territory consisting of patches of landscape, that can be either vacant or occupied by any of the species. By ``occupied'' we mean that some individuals have their home range, or their territory, in the patch, and that they keep it there for some time. This occupation can be a single individual, a family, a herd, etc. Our scale of description does not distinguish these cases, just the occupation of the patch. In order to account for human or environmental perturbations that may render parts of the landscape inhabitable, we consider that a fraction $D$ of the patches are \emph{destroyed} and not available for occupation~\cite{tilman94,bascompte96}. 

The dynamics of occupation and abandoning of patches obeys the different ecological processes that drive the metapopulation dynamics. Vacant patches can be colonized and occupied ones can be freed, as will be described below. Predators can only colonize patches already occupied by prey (as in~\cite{swihart01}, see also \cite{srivastava2008}). Predation will be taken into account as an increase in the probability of local extinction of the population of the herbivores in the presence of a local population of predators~\cite{swihart01}. 

Also, as in~\cite{tilman94}, we consider that the two competing herbivores are not equivalent. In this scenario, the superior competitor can colonize any patch that is neither destroyed nor already occupied by themselves, and even displace the inferior one when doing so. On the other hand, the inferior competitor can only colonize patches that are neither destroyed, nor occupied by themselves, \emph{nor} occupied by the superior one.

As we mentioned in the Introduction, the size and temperament of guanacos allow them to displace sheep from the scarce water sources. Besides, being native to the steppe, they are better adapted to the harsh conditions of vegetation and water availability. In this spirit, we choose to model the guanaco at the higher place in the competitive hierarchy. Sheep, being the inferior competitor, need to display some advantage in order to persist under these conditions. As observed in~\cite{tilman94}, this can be implemented in their dynamical parameters: being a better colonizer, for example. This indeed happens since---as we also mentioned in the previous section---sheep have the assistance of their owners. 

\begin{figure}[t]
\centering 
\includegraphics[width=7cm, clip=true]{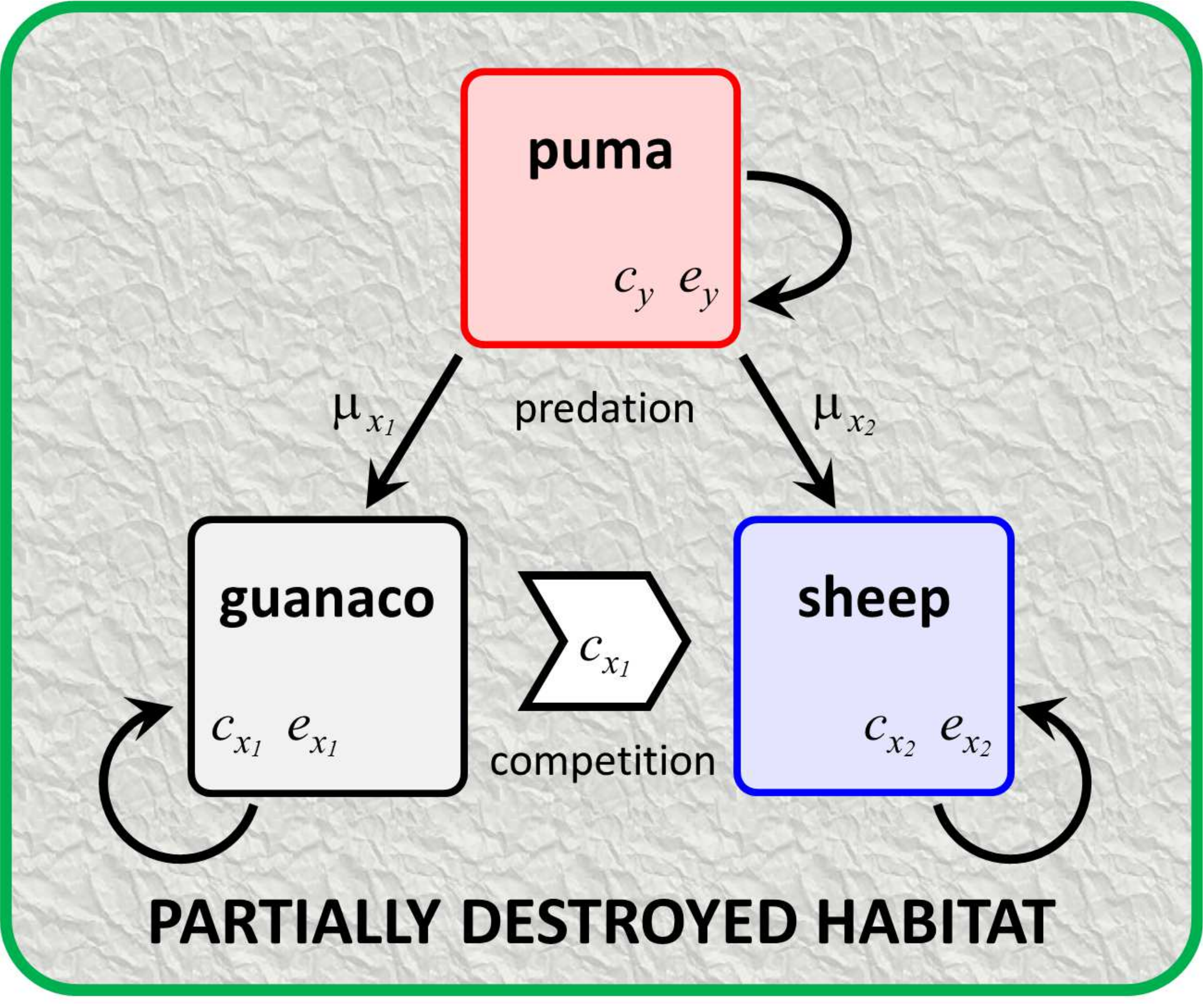}
\caption{The conceptual model. Arrows show species interactions in the model. The big arrow connecting guanacos to sheep represents the asymmetry of the hierarchical competition.}
\label{cajitas}
\end{figure}

With these conceptual predicaments, inferred from the ecosystem and represented as a diagram in Fig.~\ref{cajitas}, we can build a simple but relevant spatially extended mathematical model in the following way. Consider a square grid of $L\times L$ patches, that can be either destroyed (permanently, as a quenched element of disorder of the habitat), vacant or occupied by one or more of the species. Let $x_i$ denote the fraction of patches occupied by herbivores of species $i$ (with $i=1,\,2$ for the superior and inferior ones respectively), and $y$ the fraction occupied by predators. Time advances discretely. At each time step, the following stochastic processes can change the state of occupation of a patch:

\textbf{Colonization.} An available patch can be colonized by the species $\alpha$ from a first neighbor occupied patch, with probability of colonization $c_\alpha$ ($\alpha$ being $x_1$, $x_2$ and $y$). 

\textbf{Extinction.} An occupied patch can be vacated by species $\alpha$ with probability of local extinction $e_\alpha$. 

\textbf{Predation.} A patch that is occupied by either prey and by the predator has a probability of extinction of the prey, given by a corresponding probability $\mu_\alpha$ (note that $\mu_y=0$).

\textbf{Competitive displacement.} A patch occupied by both herbivores can be freed of the inferior one $x_2$ with probability $c_{x_1}$. Note that there is no additional parameter to characterize the hierarchy: the colonization probability of the higher competitor plays this role.

The colonization process deserves further considerations. On the one hand, lets us be more specific about the meaning of an available patch. Observe that, for the herbivores, availability is determined by the destruction, occupation and the hierarchical competition, as described above. For the predator an available patch is a patch colonized by any prey. On the other hand, colonization is the only process that involves the occupancy of neighbors besides the state of the patch itself. We must calculate the probability of colonization given the occupation of the neighbors. Let $n$ be the number of occupied neighbors of an available patch. The probability of being colonized from \emph{any} of the neighbors is then the following:
\begin{equation}
p_\alpha = 1 - (1-c_\alpha)^n.
\end{equation}

We study the dynamics of this model through a computer simulation performed on a system enclosed by impenetrable barriers (effectively implemented by destroying all the patches in the perimeter). To perform a typical realization we define the parameters of the model and destroy a fraction $D$ of patches, which will not be available for colonization for the whole run. Then, we set an initial condition occupying at random 50\% of the  available patches for each herbivore species. Further, 50\% of the patches occupied by any herbivore are occupied by predators.  The system is then allowed to evolve synchronously according to the stochastic rules. Each patch is subject to the four events in the order given above. A transient time (typically 3000 time steps) elapses before a fluctuating steady state is reached, where we perform our measurement as temporal and ensemble averages of the space occupied by each species. 

\subsection{Predation control and habitat destruction}

We have analyzed this system in several scenarios, corresponding to different values of the parameters. In particular, we show below the dependence of the state of the system on the probability of extinction of predators and on the degree of destruction of the habitat, both of which represent typical mechanisms of anthropogenic origin that affect the populations.

Let us first analyze the fraction of occupied patches of the three species as a function of the extinction probability of the predator, $e_y$. The probabilities characterizing the rest of the processes have been chosen to represent the realistic scenarios discussed above, with the guanaco as the superior competitor, while the sheep, inferior in the competitive hierarchy, is able to survive thanks to a higher colonization probability, $c_{x_2} > c_{x_1}$ and lower extinction, $e_{x_2}< e_{x_1}$. Besides, we have modeled the predation on sheep as more frequent than that on guanaco, with $\mu_{x_2} \gg \mu_{x_1}$. The colonization probability of the predators was chosen in order to have a region of coexistence of the three species.

\begin{figure}[t]
\begin{center}
\includegraphics[width=8cm,clip=true]{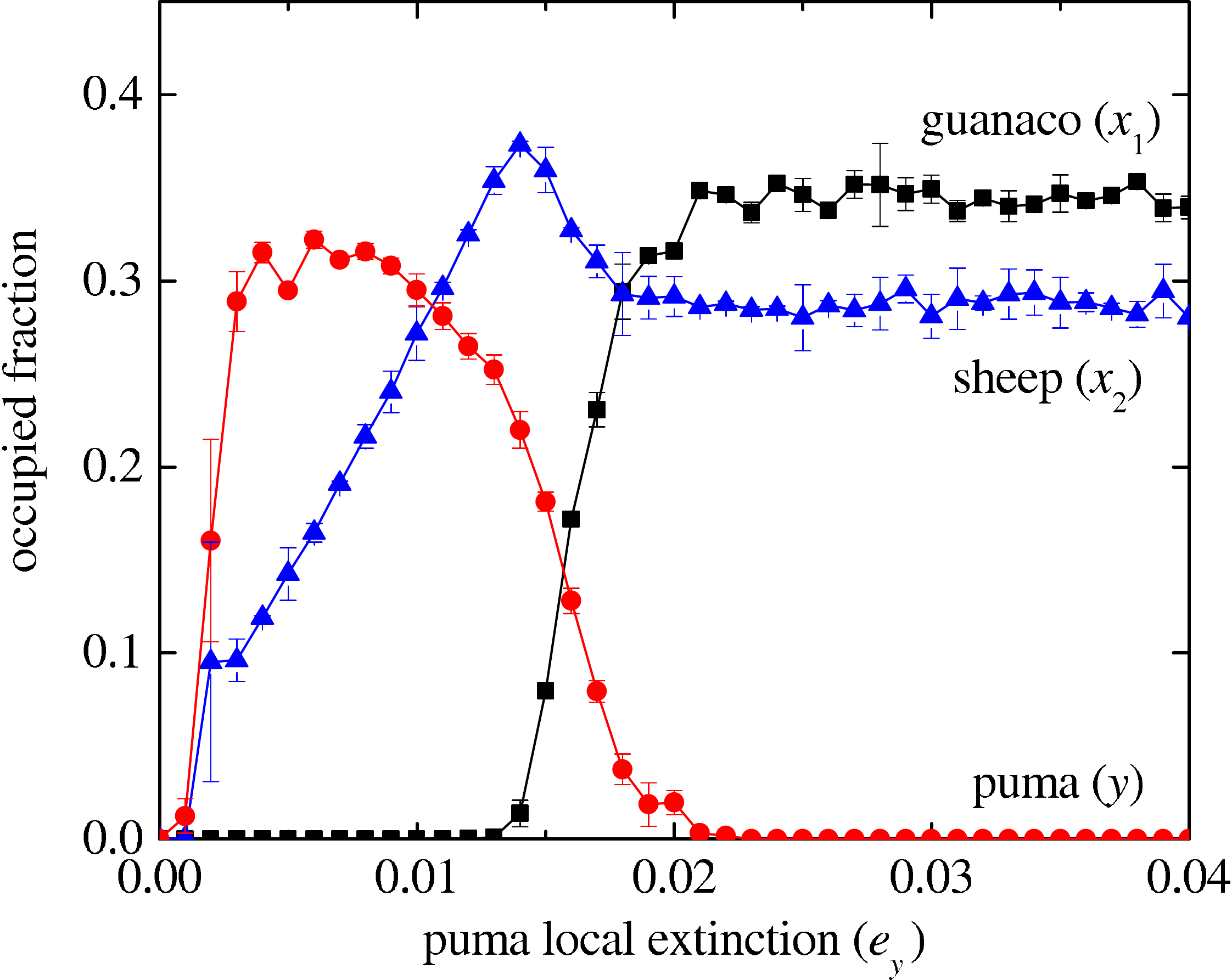}
\end{center}
\caption{Controlling the predator. We plot the fraction of occupied patches of the three species, as a function of the extinction probability of the predator. Each point is the average of the steady state, measured at the final 2000 steps, and averaged over 100 realizations of the dynamics (standard deviation shown as error bars). Other parameters are: $c_{x_1}=0.05$, $c_{x_2}=0.1$, $c_{y}=0.015$, $e_{x_1}=0.05$, $e_{x_2}=0.01$, $\mu_{x_1}=0.2$, $\mu_{x_2}=0.8$ and $D=0.3$}
\label{d3-rnd}
\end{figure}

Figure~\ref{d3-rnd} shows such a case, for a fraction $D=0.3$ of destroyed patches uniformly distributed at random in the grid. We observe several regimes of occupation of space. On the one hand, as expected, for $e_y$ above a threshold (which in this case is $\approx 0.02$), the puma becomes extinct and both herbivores coexist occupying a fraction of the system around 30\%. In this case we observe that $x_1 > x_2$, but the situation actually depends on $D$ since the \emph{superior} competitor is more sensitive to the destruction of habitat (also observed in~\cite{tilman94,bascompte96}). Other values of $D$ are reported below. 

On the other hand, the behavior of the system for very small values of $e_y$ is counter-intuitive. We see that, even though the extinction of the predator is very small, they nevertheless become extinct, and so do their preys. The reason for this is an ecological meltdown during a transient dynamics: the pumas rapidly fill their available habitat (because they almost do not vacate any occupied patch, since $e_y \approx 0$). As a consequence there is an excessive predation and the preys become extinct. Thereafter the predators follow the same fate. Figure S1 (left panel) in Supplementary Material shows a typical temporal evolution of this situation.

The most interesting dynamics is observed for intermediate values of the probability of extinction of the predators. We see that, as expected, the fraction occupied by pumas decays monotonically with $e_y$. The sheep, being better colonizers than guanacos, are the first to benefit from this lowering pressure from predation. We see this as a steady increase of the sheep occupation until some guanacos are able to colonize the system. When this happens, the conditions for the persistence of preys are modified. The sheep are now subject to both ecological pressures: predation and competitive displacement. As a consequence, we see a \emph{reduction} of the space occupied by sheep, accompanied by a fast growth of the fraction of patches occupied by guanacos. We show in Fig. S1 (right panel) in Supplementary Material a typical temporal evolution of the occupations in a situation of coexistence of the three species.

\begin{figure}[t]
\begin{center}
\includegraphics[width=\columnwidth]{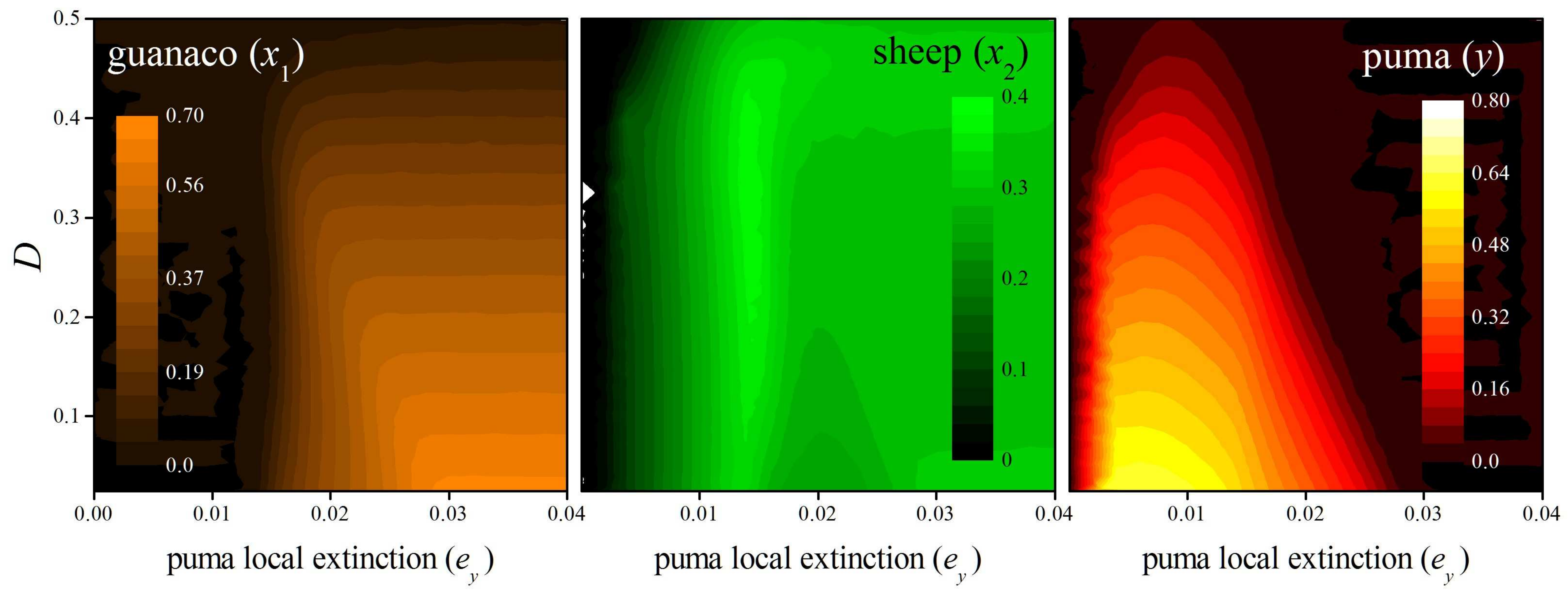}
\end{center}
\caption{Role of the habitat destruction. Contour plots of the fraction of patches occupied by each species, corresponding to parameters $e_y$ and $D$ as shown in the axes titles. (left) guanacos, $x_1$; (center) sheep, $x_2$; (right) pumas, $y$. Other parameters as in Fig.~\ref{d3-rnd}.}
\label{contour}
\end{figure}

The available area, or the fraction of destroyed habitat, also plays a relevant role in the final state of the system. Figure~\ref{contour} shows contour plots corresponding to the three species in the parameters space defined by $e_y$ and $D$. Note that the case just discussed of Fig.~\ref{d3-rnd} corresponds to a horizontal section of each of these graphs. The behavior of the space occupied by guanacos is monotonic both in $e_y$ and $D$. The threshold above which there is a non-zero occupation is independent of $D$, while the value of $e_y$ above which the occupation reaches a plateau reduces linearly with $D$. At the same time, the value of such occupation is smaller. This is the same effect on the superior competitor under the destruction of habitat, as mentioned above. 

The most prominent feature of the space occupied by sheep is the existence of a ridge (seen as an island of brighter green shades in the plot) at intermediate values of $e_y$, with a local maximum at a certain value of the destruction $D$. This is also a consequence of the hierarchical competition between sheep and guanacos. There is also a local minimum around $e_y = 0.02$ for small values of $D$. 

In the rightmost panel of Fig.~\ref{contour} we see that the extinction of pumas for very low values of $e_y$ observed in Fig.~\ref{d3-rnd} persists for all values of $D$. Observe that the threshold for extinction moves towards smaller values of the probability $e_y$ when $D$ grows. This is consistent with the known observations of greater sensitivity of predators to habitat fragmentation~\cite{srivastava2008}. Besides, the maximum fraction of occupation is independent of $e_y$ and decreases with $D$. As will be shown below, this situation changes when the distribution of destroyed patches is not uniform in the system.

In the Supplementary Material we provide three figures corresponding to the contour plots of Fig.~\ref{contour}, showing horizontal cuts for several values of $D$. These results will be further discussed in the final Section.

\subsection{Mean field approximation}

Let us briefly discuss a mean field approximation of the spatially extended model, which is similar to the original Levins model of metapopulations and subsequent generalizations~\cite{melian02,swihart01,harding2002,bascompte98,roy2008}. It is important to stress that such a model ignores the short-range correlations between occupied patches that arise from the local and first-neighbor population dynamics. These correlations, as we will see in the next Section, play an important role even in the average behavior of the space occupation. For this reason, we regard the numerical simulation of the model as a more valuable tool for the analysis of the system. Nevertheless, some global features of the metapopulations can still be captured by an analytical model and it is worth some consideration for the insight it provides in the mechanisms behind the observed phenomena.

Taking into account the four processes described above, the Levins-like mean field approximation for the three-species model becomes:
\begin{eqnarray}
\frac{dx_1}{dt} &=& c_{x_1} x_1 (1-D-x_1)     - e_{x_1} x_1 - \mu_1 x_1 y, \label{mfx1}\\
\frac{dx_2}{dt} &=& c_{x_2} x_2 (1-D-x_1-x_2) - e_{x_2} x_2 - \mu_2 x_2 y -c_{x_1} x_1 x_2, \label{mfx2}\\
\frac{dy}  {dt} &=& c_{y} y \left[(x_1+x_2-x_1 x_2)-y\right] - e_y y, \label{mfy}
\end{eqnarray}
where $c_\alpha$, $e_\alpha$ and $\mu_i$ are colonization, extinction and predation rates of the corresponding species. Observe the difference in the colonization terms between the two competing herbivores, Eqs.~(\ref{mfx1}) and (\ref{mfx2}). As mentioned above (and as in~\cite{tilman94}) the patches available for colonization by the higher competitor in the hierarchy, $x_1$, are those not destroyed and not already occupied: ($1-D-x_1$). The lower competitor $x_2$, instead, can colonize patches which, besides, are not occupied by $x_1$: $(1-D-x_1-x_2)$. The last term of Eq.~(\ref{mfx2}), in addition, stands for a displacement mechanism of the lower competitor by the higher one. In this situation the lower competitor can persist only for a certain set of parameters, granting a better colonization or less extinction rates. Observe, also, that the colonization term of the predator (Eq.~(\ref{mfy})) allows for them only the occupation of patches already occupied by either prey (similarly to the proposal of~\cite{bascompte98}). The form of this term takes care of this interference effect following the rule of addition of probabilities for the union of the patches already colonized by $x_1$ and $x_2$. Finally, the predation terms present in Eqs.~(\ref{mfx1}) and (\ref{mfx2}) are proportional to the fraction of space occupied by predators which, even in this mean field without any space dependence, corresponds to a high mobility of predators in the system. 

\begin{figure}[t]
\centering 
\includegraphics[width=\columnwidth, clip=true]{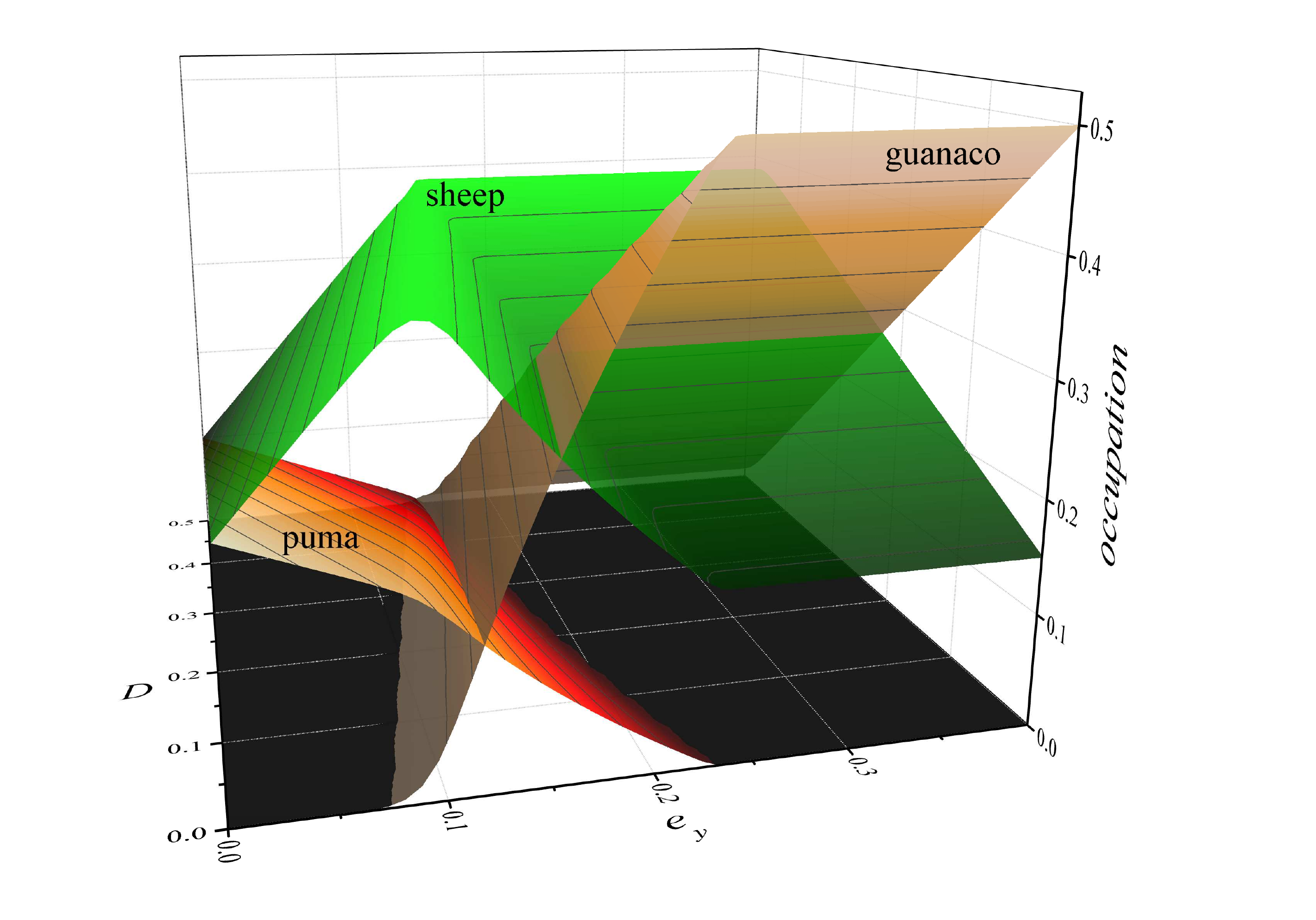}
\caption{Phase diagram of the mean field model. The plot shows the stationary occupation of patches of the three species as controlled by the destruction parameter $D$ and the extinction of predators $e_y$. }
\label{mfphasediag}
\end{figure}

The phase diagram of Fig.~\ref{mfphasediag} shows the equilibrium state of the system (\ref{mfx1})-(\ref{mfy}), i.e. the steady state of the occupied fractions of the three species. It corresponds to the scenarios shown in Fig.~\ref{contour}. It is apparent that the behavior of the mean field model is qualitatively similar to the results obtained in the spatially extended computer simulations. In particular, the non-monotonic dependence of the fraction of patches used by sheep with the extinction rate of pumas stands out as a robust signature of this three-species food web. We will discuss some implications of this feature in the Discussion.

\begin{figure}[t]
\centering 
\includegraphics[width=\columnwidth]{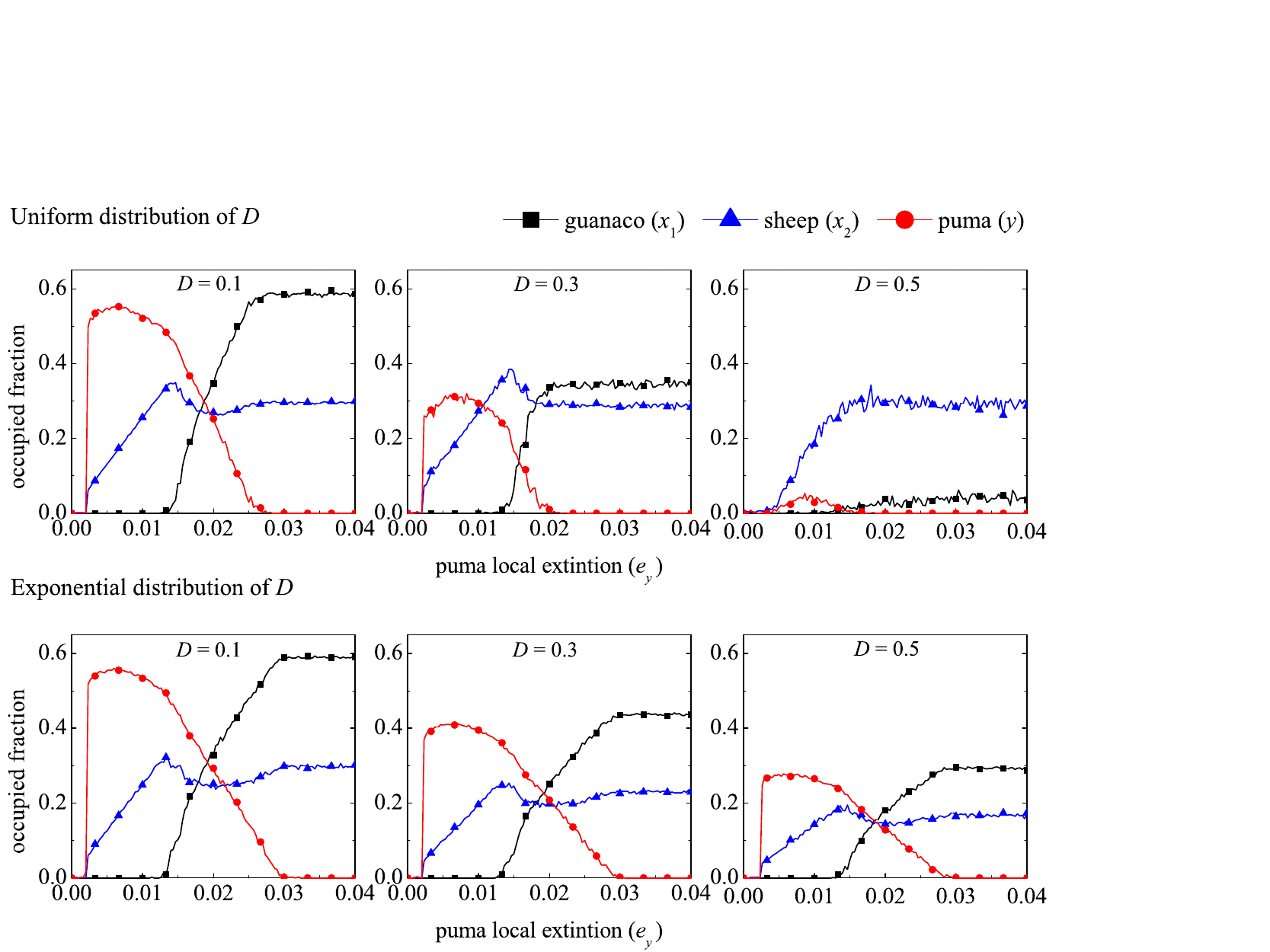}
\caption{Non-uniform destruction of the habitat. Comparison of the results of simulations performed with the same values of $D$ (as shown), but with either a uniform (top) or an exponential distribution in one direction of the grid (bottom). Only the average values are shown for clarity, since the standard deviations are very small, as those shown in Fig.~\ref{d3-rnd}. Each curve has 120 points (of which only a few are shown with symbols for clarity) and each point is the average of 100 realizations.}
\label{unif-vs-exp}
\end{figure}

\subsection{Limitations of the mean field approximation}

As anticipated, a simple mean field approximation cannot take into account peculiarities of the dynamics that arise from spatial correlations of the variables. Of particular relevance in the present context is  the distribution of the destroyed patches. The results shown above correspond to a fraction $D$ of destroyed patches distributed at random in a uniform manner on the grid. However, in many situations the distribution of destroyed patches may be non-uniform, albeit also random. It may follow a gradient, corresponding to an analogous distribution of resources. Or it may represent some spatially organized human activities such as agriculture, urban development, roads, etc. Eventually, one may also consider the feedback of the variables of the system on the state of the habitat, as in the case of desertification due to overgrazing---a phenomenon that we will explore elsewhere. 

We have analyzed several instances in which the aggregation of destroyed patches affects the state of the system, and we show a simple one here. Consider that the probability distribution of the $D\times L^2$ destroyed patches is random in one direction of the grid, and decays exponentially in the other direction. (See, in Fig. S3 of the Supplementary Material, two examples of the uniform and exponential distributions of destroyed habitat.) With such a distribution, there is a higher local density of destroyed patches near one of the sides of the system, decaying towards the opposite side. Characteristic results are shown in Fig.~\ref{unif-vs-exp}, which compares both uniform and exponential distributions of destruction for the same three values of $D$, corresponding to low destruction (left column), moderate destruction (center) and high destruction of the habitat (right).

We see that the main difference between the two arrangements is the response of the pumas and guanacos. Both of them seem very sensitive to the spatial arrangement of unusable patches. The reason for this is not completely clear, but it seems to arise from the fact that the predators, since they occupy a subset of the patches colonized by preys, have a more limited chance of survival than the preys. On the other hand the guanacos that, being the superior competitor are more susceptible to the destruction of habitat (as observed by~\cite{tilman94}), suffer a similar fate. When the destroyed patches are uniformly distributed in the system they impose a rather strong handicap for colonization when the value of $D$ is high (indeed, $D=0.5$ is subcritical for the percolation of undestroyed patches~\cite{bunde91}). The top row of Fig.~\ref{unif-vs-exp}, from left to right, corresponds to a progressive reduction of the occupied space as well as $D$ increases. Eventually, for $D=0.5$ we see that the pumas and the guanacos have almost disappeared. On the other hand, the bottom row of plots in the same figure shows that the exponential distribution of the destruction has a much smaller impact on the space occupied by these species. Indeed, such distribution concentrates the unusable space near one of the sides of the system, leaving relatively pristine the opposite one. This enables the colonization of this range with less local hindrance, as if they were effectively in a smaller  system with a lower value of $D$. Moreover, the threshold of $e_y$ for the extinction of pumas becomes independent of $D$ in this heterogeneous system.

To provide further support to this argument we analyzed another distribution of the destroyed patches. When they are placed as a single block of unusable space (effectively reducing the available area to a contiguous set of $(1-D)L^2$ patches), the result is both quali- and quantitatively very similar to the one presented here of an heterogeneous habitat with an exponentially distribution of the destruction.

\section{Discussion}

The results presented in the previous Sections are a sample of the rich phenomenology of the model. There are a number of aspects that deserve further discussion, and we delve into some of them in this Section. On the one hand, consider that a thorough exploration of the dynamics of a model with three species and nine parameters is a daunting task. For this reason, we have chosen to restrict our analysis to a small region in parameter space, bearing in mind the correspondence to realistic natural phenomena. Nevertheless, concerned by the robustness of our results, we extended our study to a wider region of parameters around those reported here. We found no qualitative departure from the shown results. We also explored the structural stability of the model by incorporating certain additional mechanisms, namely an exploration cost for predators capable of resource supplementation (as in~\cite{swihart01,melian02}), with the same encouraging results. 

\begin{table}[t]
\centering
\begin{tabular}{|l|c|c|c|}
\cline{1-4}
Species        & Low conflict           & Medium conflict        & High conflict           \\ \cline{1-4}
               & $D=0.1$                & $D=0.3$                & $D=0.5$                 \\ \cline{1-4}
guanaco ($x_1$)& $0.35\rightarrow 0.47$ & $0.32\rightarrow 0.45$ &  $0.03\rightarrow 0.20$ \\ \cline{1-4}
sheep ($x_2$)  & $0.26\rightarrow 0.00$ & $0.29\rightarrow 0.00$ & $0.29\rightarrow 0.00$  \\ \cline{1-4}
puma ($y$)     & $0.25\rightarrow 0.48$ & $0.02\rightarrow 0.17$ &  $0.00\rightarrow < 0.01$ \\ \cline{1-4}
\end{tabular}
\caption{Modeling of the conflict scenarios. The occupied fractions reported for each species correspond to the stationary state of the model. The arrows indicate the change of this asymptotic state under the remotion of the sheep and the corresponding change of the parameters as: $c_{x_1}=0.05\rightarrow 0.1$, $e_{x_1}=0.05\rightarrow 0.025$, $e_y=0.02\rightarrow 0.015$, $\mu_{x_1}=0.2\rightarrow 0.3$. Other parameters as in Fig.~\ref{d3-rnd}. The habitat has uniform distribution of destroyed patches.}
\label{tabla}
\end{table}

Let us revisit the three conflict scenarios discussed in the Introduction. Within the limitations of the present model it is possible to assess some of their features. The following considerations are summarized in Table 1. Consider first the low conflict scenario, that we can visualize as a coexistence of the three species in a situation of little overgrazing. Such a situation could correspond to the set of parameters represented in the top-left panel of Fig.~\ref{unif-vs-exp}, with $e_y=0.02$ and coexistence of the three species. What would change if the use of the land changes from productive to conservative? The first change would be the elimination of the livestock, i.e. $x_2=0$. This would be accompanied by a number of changes in several parameters (see the caption of Table 1). The disappearance of the ``bodyguards'' of the sheep would correspond to an increase in the colonization rate of guanacos, and a decrease of the extinction rates of guanacos and pumas. The predation pressure on guanacos would also increase due to the lack of sheep. Observe in the first column of Table 1 the change that such a scenario produces in the model: a sharp recovery of the wildlife. The same arguments can be applied to the medium and the high conflict scenarios, as characterized by increasing values of $D$ (see the center and rightmost panels of Fig.~\ref{unif-vs-exp}). The medium conflict scenario shows a similar recovery of the guanacos than the low conflict one, but at a much slower pace (not reported in the table). It is also remarkable that the high conflict scenario shows just a little recovery of the guanacos, with almost no effect in the puma population. Actually, the removal of sheep would improve the available space by, eventually, reducing $D$. The treatment of the parameter $D$ as a dynamical variable of the system will indeed be explored in the future, as will the existence of independent resources for the various species.

In the Supplementary Material, Fig. S4 shows typical temporal evolutions corresponding to the three scenarios just discussed.

\begin{figure}[t]
\centering 
\includegraphics[width=\columnwidth, clip=true]{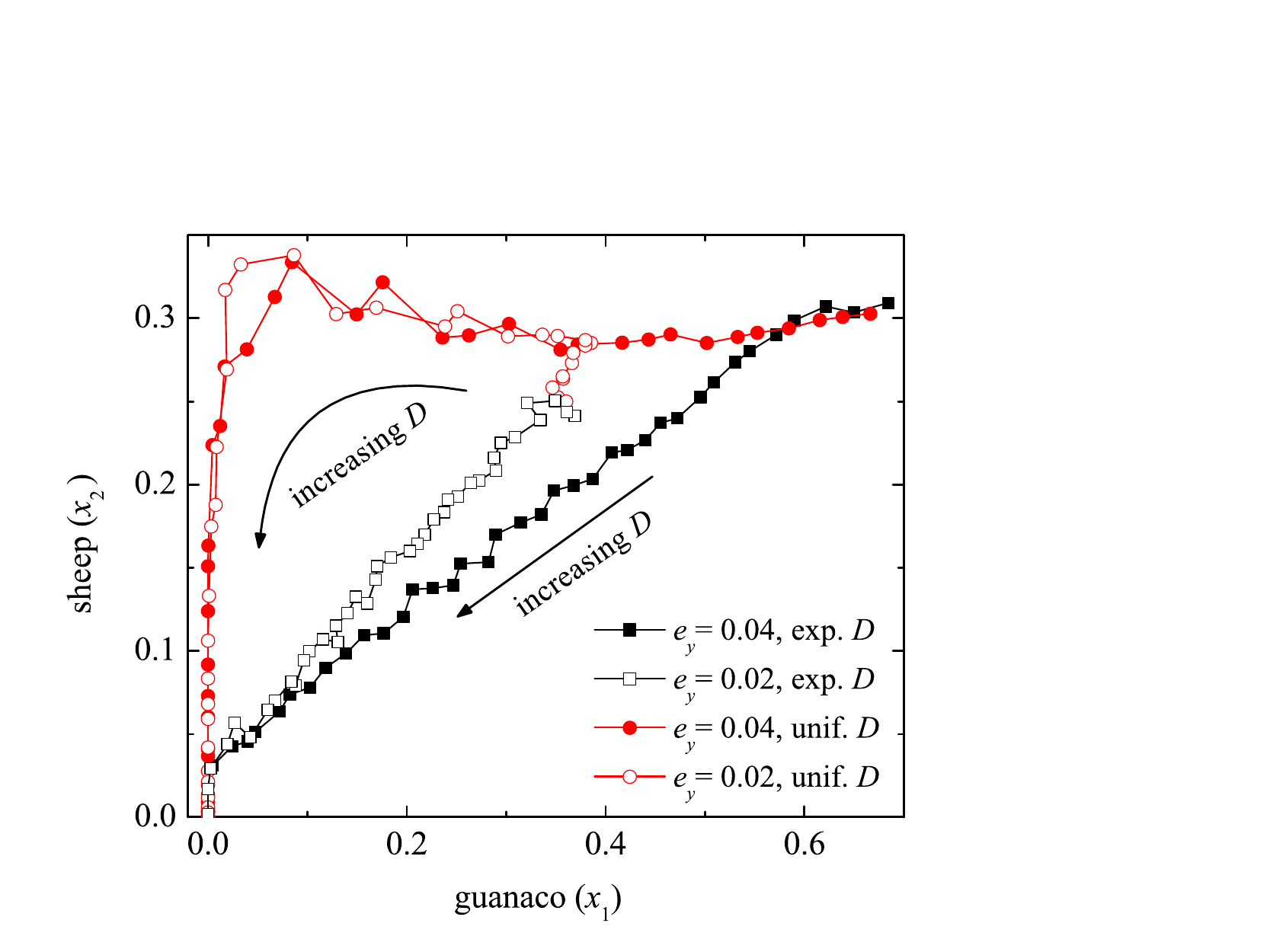}
\caption{Meltdown of the herbivores. The curves show the trajectory of the occupied fraction in the space defined by the two herbivores, when the patches are progressively destroyed (as indicated by the arrows, from $D=0$ to $D=1$). Each curve corresponds to different distributions of the destroyed patches and probability of extinction of the predator, as shown. Other parameters as in Fig.~\ref{d3-rnd}.}
\label{meltdown}
\end{figure}

It is also worth noting the role of the predator as a keystone species. Paine~\cite{paine1969} introduced the concept to characterize the existence of species that, despite their relative low abundance in an ecosystem, play a highly critical role in the ecological dynamics. The ecosystem may suffer a drastic change if a keystone species is removed. In most of the cases the keystone species is a predator that can control the distribution and population of large numbers of prey species~\cite{mills1993}. Such cases of top-down control of the ecosystem have even been the subject of field experimentation~\cite{terborgh2001}. The behavior of the populations of sheep and guanaco as a function of the abundance of puma may indicate that the latter is acting like a keystone predator. The change in prevalence of the herbivore species due to the extinction of the puma has been both verified in the mean field model and in numerical simulations.

As a final comment, let us insist on the importance of the \emph{distribution} of the destroyed patches on its influence on the dynamics. It is known that certain modes of destruction of habitat, such as fragmentation, are more deleterious than others. In our model we have observed this fact, as discussed in the relative effect of uniform or exponential distribution of the destroyed patches (Fig.~\ref{unif-vs-exp}). A clearer way of representing this effect is shown in Fig.~\ref{meltdown}. Here we plot the final state of the system in the plane defined by $x_1$ and $x_2$. Each point corresponds to a different value of $D$, with the rest of the parameters as shown. Two distinct behaviors are observed: one corresponding to the uniform distribution and another one to the exponential one. Besides, the different symbols (solid or open) correspond to very different situations: the solid ones have the puma population completely extinct, while the open ones have three-species coexistence. The value of $D$ increases as indicated by the arrows. In the exponential case both species get proportionally reduced with the increase of $D$, until the guanacos disappear with a small residual population of sheep. The presence of pumas in the system can be recognized by a different slope of this straight line: the predators accelerates the extinction of sheep, their main prey. On the other hand, the uniform case follows a different course. The presence of pumas is not relevant in the meltdown of the system. Besides, the two herbivores are very differently affected. The guanacos are more strongly affected and disappear earlier, and get extinct at $D\approx 0.6$ (see Fig. S5 in the Supplementary Material). This, indeed, is the critical value of percolation of destroyed patches on a square lattice~\cite{bunde91}. The reason for this is, again, the relative advantage of sheep that, being the inferior competitor, need a better colonization rate. This improves their chances of thriving in a space that gets very fragmented and disconnected as the destruction progresses.

In many situations of conflict between production and conservation, it is difficult to make ecologically right decisions in terms of sustainability if the variables are taken separately, ignoring their interactions. We believe that our results highlight the importance of mathematical models in the decision-making process, and constitute a valuable contribution for the theoretical framework of interacting metapopulations.  

\section{Acknowledgements}
We acknowledge financial support from: CONICET (PIP 112-201101-00310), Universidad Nacional de Cuyo (06/C410), ANPCyT (PICT-2011-0790). The funding agencies were not involved in the research or the preparation of the manuscript.

\bibliographystyle{elsarticle-num} 
\bibliography{ganado}



\section*{Supplementary material}

# Supplementary Material for

# Mathematical model of livestock and wildlife: Predation and competition under environmental disturbances

M. F. Laguna[1,*], G. Abramson[1,2,*], M. N. Kuperman[1,2,*], J. L. Lanata[3,*] and J. A. Monjeau[4,5,*]

1 Centro Atómico Bariloche and CONICET, R8402AGP Bariloche, Argentina
2 Instituto Balseiro, R8402AGP Bariloche, Argentina
3 Instituto de Investigaciones en Diversidad Cultural y Procesos de Cambio, CONICET, R8400AHL Bariloche, Argentina
4 Fundación Bariloche and CONICET, R8402AGP Bariloche, Argentina
5 Laboratório de Ecologia e Conservação de Populações, Departamento de Ecologia, Universidade Federal do Rio de Janeiro, Rio de Janeiro, Brasil

## Section 1. Temporal evolution

In this section we provide details of the dynamical behavior of the system. We show in Fig. S1 two single runs corresponding to the parameters of Fig. 2 of the main paper, for the cases $e_y=0.001$ and $0.017$. The left panel corresponds to a negligible extinction of the predator, which induces a meltdown of the system. This situation is probably unobservable in any real system due to the unrealistic small value of the mortality of the predator. The right panel shows a situation in which the three species coexist.

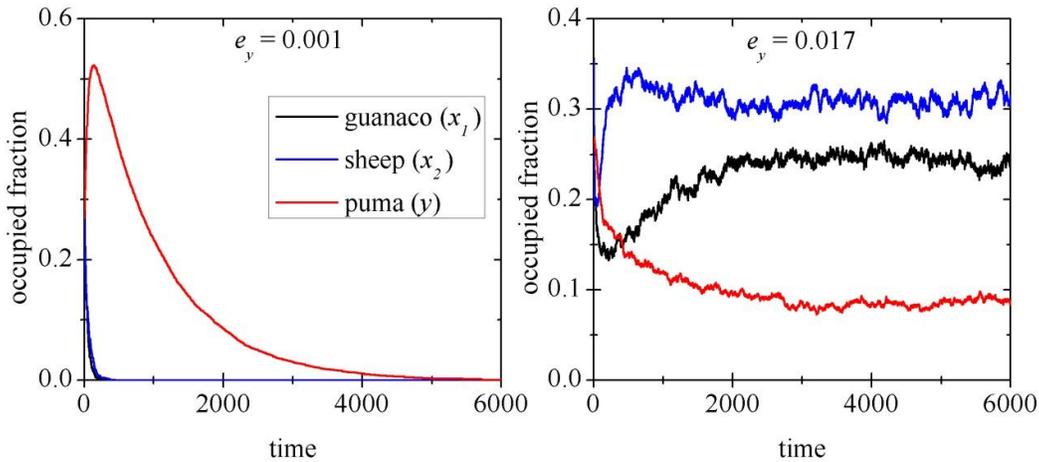

**Figure S1.** Fraction of occupied patches for the three species as a function of time. The value of the puma local extinction $e_y$ for each case is indicated in the panels. Other parameters as in Fig. 2 of the main paper.

## Section 2. The role of the habitat destruction

Figure S2 shows three plots which complement the Fig. 3 of the main paper. Each curve is a horizontal cut of that contour plot, corresponding to different values of the parameter $D$. For the guanacos (left panel), note that the threshold of extinction is independent of $D$. In the central panel, observe the local maximum of occupation by sheep in the curves, corresponding to the

---
*E-mail addresses: lagunaf@cab.cnea.gov.ar, abramson@cab.cnea.gov.ar, kuperman@cab.cnea.gov.ar, jllanata@conicet.gov.ar, amonjeau@fundacionbariloche.org.ar

ridge mentioned in the discussion of Fig. 3. For the pumas (rightmost panel), note the maximum of the curves at the same value of $e_y$, and decreasing with $D$. Also, note that this specie has a threshold for extinction has a threshold which decreases with the increase of $D$.

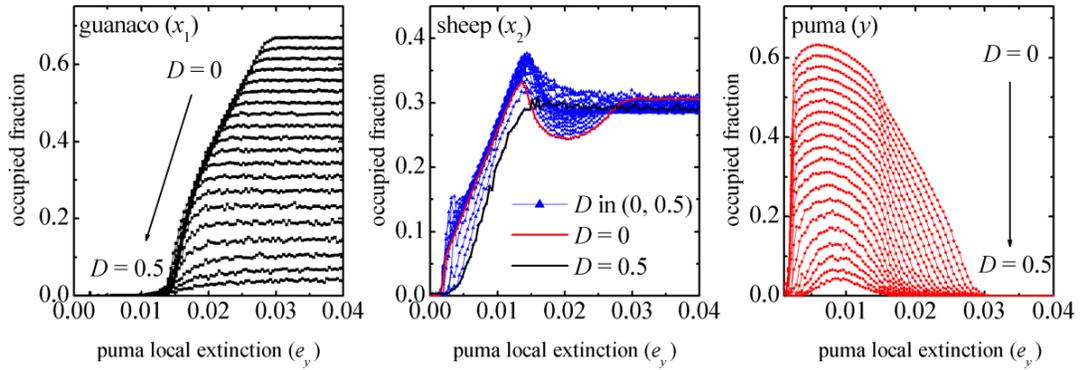

**Figure S2.** Dependence of occupied fractions on $D$ and $e_y$, for the three species. Other parameters as in Fig. 3 of the main paper.

### Section 3. Spatial distribution of destruction

As we discuss in Section II D of the main paper, the spatial distribution of the destroyed patches plays a relevant role in the species survival. Here we plot two examples of how the habitat is simulated, for the case of D=0.3. Left panel correspond to a uniform distribution of $D$ whereas right panel shows an exponential distribution of $D$.

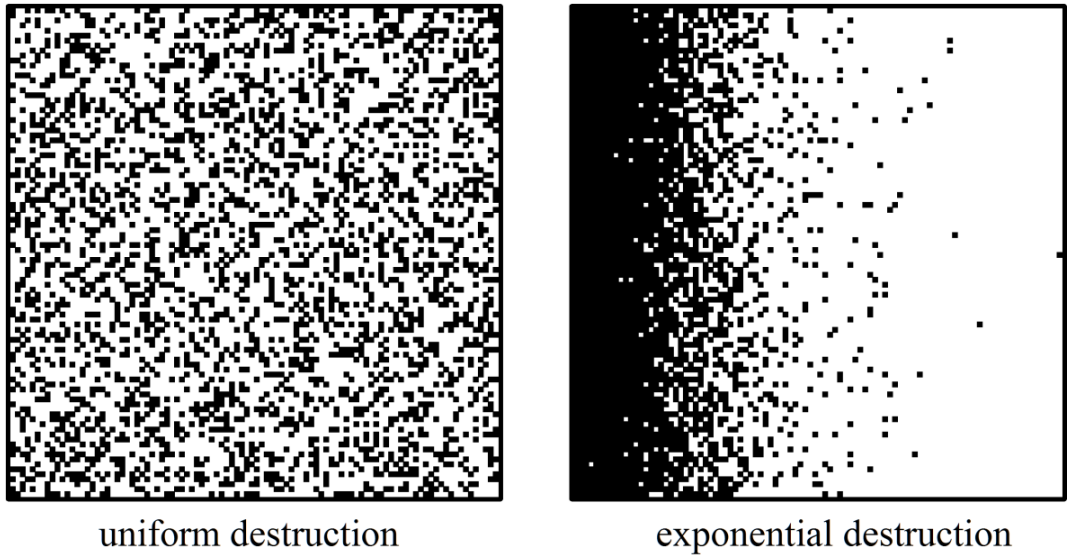

**Figure S3.** Examples of partially destroyed habitats, for $D = 0.3$. The system is a lattice of 100×100 patches. Black ones are destroyed.

### Section 4. Conflict scenarios

In this section we show the evolution of the three species for the different conflict scenarios presented in Table 1 of the main paper. Left panel shows the situation for a low conflict scenario, modelled as a system with a low level of destruction ($D$=0.1). Medium and high

conflict scenarios correspond to $D=0.3$ and $D=0.5$ respectively. In all the cases, the three species coexist during the first 5000 time steps. At time $t=5000$, all sheep are removed from the system (i.e., $x_2$ is set to 0) and the parameters are changed as indicated in Table 1. In the first to cases the native species are recovered, whereas in the high conflict case the pumas remain absent. In a real situation, a repopulation might occur by migration from more distant regions, not considered in this model. Note that the medium conflict scenario needs twice the temporal steps to reach the stationary state compared with the low conflict case.

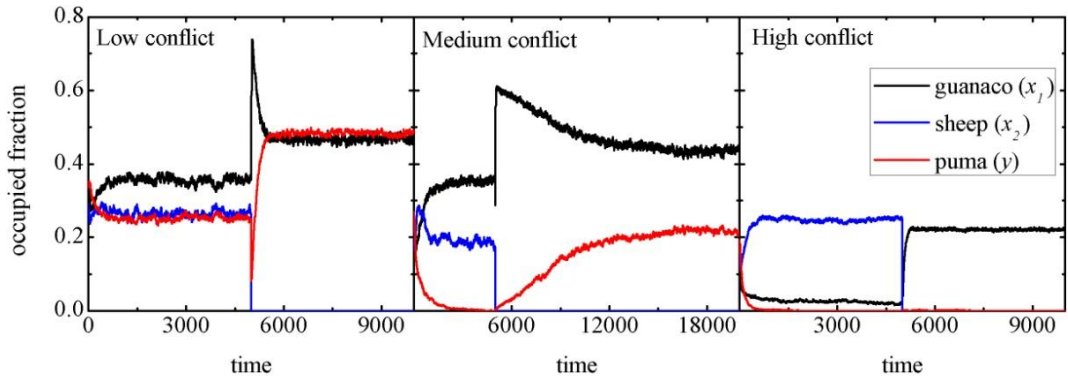

**Figure S4.** Temporal evolution of the three species, for three conflict scenarios as described in the text. At $t = 5000$ the removal of sheep is clearly seen.

### Section 5. Ecological meltdown

Figure S5 is an alternative reading of the meltdown of the herbivores discussed around Fig. 6 of the main text. Here we show results for a single value of $e_y=0.04$ (the case where the pumas are already extinct). Observe the decaying occupations as $D$ grows. Furthermore, observe that the exponential distribution of patches permits the persistence of both species at higher destruction values.

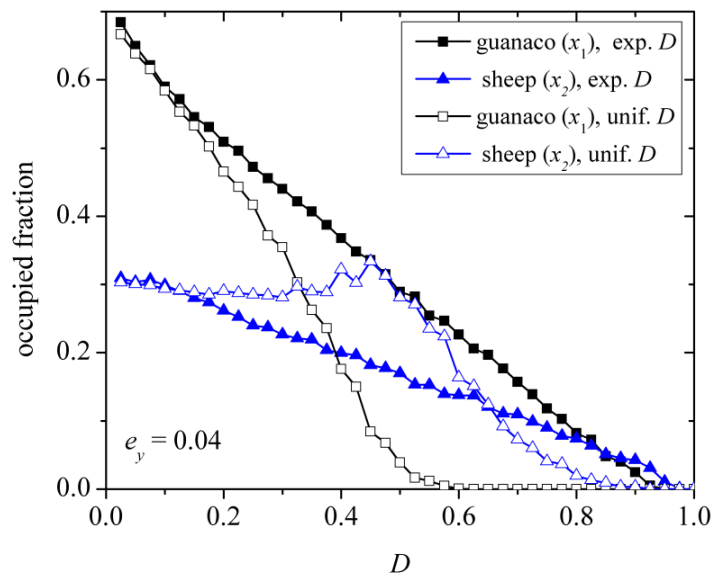

**Figure S5.** Meltdown of the herbivores. Occupied fraction of patches as a function of the fraction of destroyed habitat. Other parameters as in Fig. 2 of the main paper.


\end{document}